\documentclass[10pt, conference]{IEEEtran}
\IEEEoverridecommandlockouts
\usepackage{cite}
\usepackage{amsmath,amssymb,amsfonts}
\usepackage{mathtools}
\usepackage{bbm}
\usepackage{algorithmic}
\usepackage{graphicx}
\usepackage{textcomp}
\usepackage[table]{xcolor}
\usepackage{multirow}
\usepackage{hhline,colortbl}
\usepackage{soul}
\usepackage{siunitx}
\usepackage{subcaption}
\usepackage[printonlyused]{acronym}
\usepackage{tikz}
\usepackage{textcomp}
\usepackage{hyperref}

\acrodef{mmW}{millimeter-wave}
\acrodef{sub-THz}{Sub-terahertz}
\acrodef{BS}{base station}
\acrodef{UE}{User equipment}
\acrodef{NR}{New Radio}
\acrodef{TX}{transmitter}
\acrodef{RX}{receiver}
\acrodef{AWV}{antenna weight vector}
\acrodef{LOS}{line-of-sight}
\acrodef{NLOS}{non-line-of-sight}
\acrodef{AoA}{angles-of-arrival}
\acrodef{AoD}{angle-of-departure}
\acrodef{DA}{digital array architecture}
\acrodef{SA}{sub-array hybrid architecture}
\acrodef{FH}{fully connected hybrid architecture}
\acrodef{MU}{multi-user}
\acrodef{MIMO}{multiple-input multiple-output}
\acrodef{MU-MIMO}{multi-user multiple-input multiple-output}
\acrodef{OFDM}{orthogonal frequency division multiplexing}

\acrodef{ADC}{analog-to-digital converters}
\acrodef{LO}{Local Oscillators}
\acrodef{PS}{Phase Shifters}
\acrodef{LNA}{Low Noise Amplifiers}
\acrodef{DSP}{digital signal processing}
\acrodef{BB}{baseband}
\acrodef{VGA}{Variable Gain Amplifier}
\acrodef{FoM}{figure-of-merit}
\acrodef{NF}{noise figure}
\acrodef{IL}{insertion loss}

\acrodef{EE}{energy efficiency}
\acrodef{SE}{spectral efficiency}
\acrodef{SINR}{signal-to-interference-and-noise ratio}
\acrodef{MC}{Monte Carlo}

\newcommand\copyrighttext{%
  \footnotesize \textcopyright 2022 IEEE. Personal use of this material is permitted.
  Permission from IEEE must be obtained for all other uses, in any current or future
  media, including reprinting/republishing this material for advertising or promotional
  purposes, creating new collective works, for resale or redistribution to servers or
  lists, or reuse of any copyrighted component of this work in other works. 
}
\newcommand\copyrightnotice{%
\begin{tikzpicture}[remember picture,overlay]
\node[anchor=south,yshift=10pt] at (current page.south) {\fbox{\parbox{\dimexpr\textwidth-\fboxsep-\fboxrule\relax}{\copyrighttext}}};
\end{tikzpicture}%
}

\newcolumntype{g}{>{\columncolor{lightgray}}c}

\DeclarePairedDelimiter{\ceil}{\lceil}{\rceil}
\newcommand{\norm}[1]{\left\lVert#1\right\rVert}

\def\BibTeX{{\rm B\kern-.05em{\sc i\kern-.025em b}\kern-.08em
    T\kern-.1667em\lower.7ex\hbox{E}\kern-.125emX}}

\begin{document}
\bstctlcite{IEEEexample:BSTcontrol}


\title{Energy Efficiency Tradeoffs for Sub-THz Multi-User MIMO Base Station Receivers\\
\thanks{This work was supported by the NSF under grant 1718742. This work was also supported in part by the ComSenTer and CONIX Research Centers, two of six centers in JUMP, a Semiconductor Research Corporation (SRC) program sponsored by DARPA.}
}

\author{\IEEEauthorblockN{Benjamin W. Domae, Christopher Chen, Danijela Cabric}
		
\IEEEauthorblockA{\textit{Electrical and Computer Engineering Department,} \\
\textit{University of California, Los Angeles}\\
bdomae@ucla.edu, cchen12@ucla.edu, danijela@ee.ucla.edu}
}

\maketitle
\copyrightnotice

\begin{abstract} \ac{sub-THz} antenna array architectures significantly impact power usage and communications capacity in \ac{MU-MIMO} systems. In this work, we compare the energy efficiency and spectral efficiency of three \ac{MU-MIMO} capable array architectures for base station receivers. We provide a sub-THz circuits power analysis, based on our review of state-of-the-art D-band and G-band components, and compare communications capabilities through wideband simulations. Our analysis reveals that digital arrays can provide the highest spectral efficiency and energy efficiency, due to the high power consumption of sub-THz active phase shifters or when SNR and system spectral efficiency requirements are high.
\end{abstract}
\IEEEpeerreviewmaketitle


\section{Introduction}
As demand for cellular data increases, wide bandwidths available at upper \ac{mmW} and \ac{sub-THz} frequencies are being proposed for future wireless networks. Studies for future 5G and 6G standards are considering D-band (110-170 GHz) and G-band (140-220 GHz) as the natural progression from \ac{mmW} systems in current 5G \ac{NR} standards. However, the power consumption of transceivers at these bands, especially those capable of spatially multiplexing multiple users at once with \ac{MU-MIMO}, is a major concern for future development. While many studies have assumed \ac{BS} power consumption is irrelevant to system performance, this perspective ignores the ongoing operational costs, stricter thermal design requirements, and the environmental concerns of higher power usage.

Few prior works have considered the power consumption or energy efficiency of \ac{sub-THz} systems. The power consumption study in \cite{Skrimponis2020_subthzMobileRx} analyzed digital and analog handheld \ac{sub-THz} \ac{RX} architectures, but analog arrays do not support \ac{MU-MIMO} capability. The authors of \cite{Halbauer2021_subthzTxRx} used a theoretical link budget to compute \ac{sub-THz} \ac{RX} and \ac{TX} power. Neither examined the impact of array architectures on communications capability.

Several publications have investigated \ac{BS} power consumption at \ac{mmW} bands. \cite{Yan2019_MCAS} detailed tradeoffs in array architecture, communications capability, power consumption, and circuit area for 28 GHz \ac{BS} \ac{TX}s but not \ac{RX}s. Additionally, the trade study in \cite{Abbas2017_mmwRx} compared the theoretical capacity and energy efficiency of 60 GHz \ac{RX} designs, but did not use simulations to evaluate realistic achievable rates.

This work investigates the energy efficiency of \ac{sub-THz} \ac{BS} \ac{RX}s with three \ac{MU-MIMO} capable array architectures. We compare the \ac{DA}, the \ac{SA}, and the \ac{FH} for power consumption and simulated communications performance. Our detailed circuits and systems modeling, utilizing a literature survey of recent D-band and G-band components, provides a higher fidelity power model than prior studies at \ac{sub-THz}. Wideband, symbol-level, \ac{MU-MIMO} \ac{OFDM} simulations then provide realistic estimates of each architecture's communications capabilities. Compared to prior work, we provide a more realistic trade study for \ac{sub-THz} \ac{BS} \ac{RX}s.


\section{Problem Statement}
\label{sec:problem}
\begin{figure*}
    \centering
    \includegraphics[width=\textwidth,trim=15 15 15 15,clip]{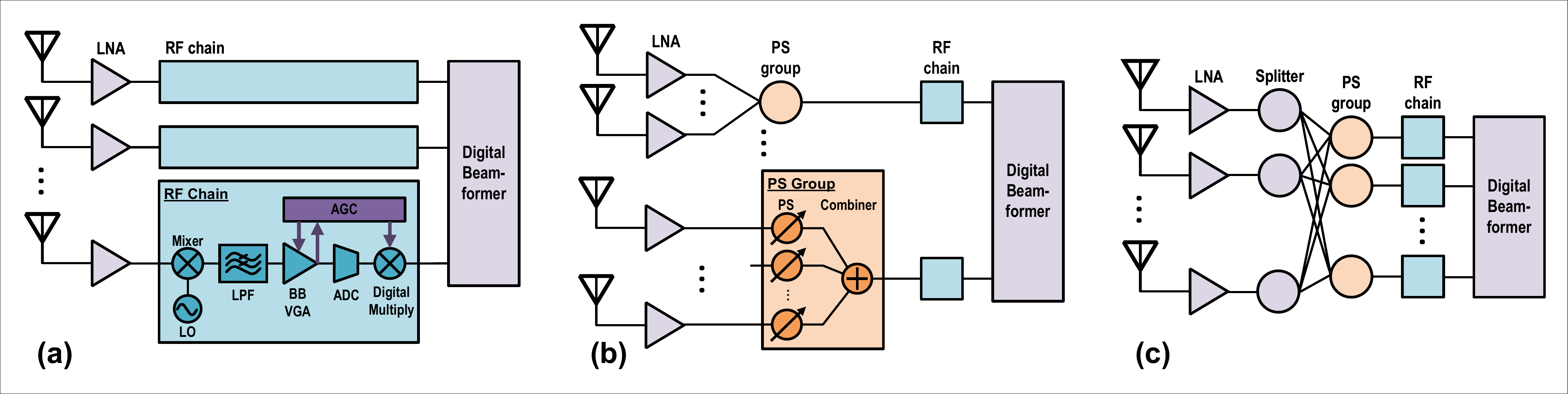}
    \caption{(a) \ac{DA}, (b) \ac{SA}, and (c) \ac{FH} \ac{RX} array architectures considered.}
    \label{fig:architectures}
    \vspace{-3mm}
\end{figure*}

To understand the \ac{RX} power consumption, we first review the three array architecture designs. Fig. \ref{fig:architectures} shows block diagrams for the \ac{DA}, \ac{SA} and \ac{FH}. Table \ref{table:components} then summarizes the total number of each component required for each design. The number of RF chains, $N_{RF}$, limits both the maximum number of simultaneous streams the \ac{RX} can support and the degrees of freedom for MIMO combining. While $N_{RF}$ for the \ac{DA} is fixed to the number of \ac{BS} antennas ($N_{BS}$), $N_{RF} < N_{BS}$ is a design parameter for the \ac{SA} and the \ac{FH}. Although this work assumes $N_{RF} = U$, where $U$ is the number of users simultaneously connected to the \ac{BS}, the \ac{SA} and \ac{FH} each offer unique combinations of hardware requirements and degrees of freedom for combining. Compared to the \ac{DA}, the \ac{SA} and \ac{FH} reduce communications capability and increase the number of RF components in order to reduce the number of baseband components. Additionally, \ac{ADC} resolution and wideband channels make the tradeoff more complex. \textbf{The goal of this work is to evaluate this tradeoff between \ac{RX} architecture, power consumption, communications capability, and \ac{ADC} resolution using realistic circuit component power}.

    

\begin{table}[htbp]
\caption{Summary of the number of required components}
\vspace{-3mm}
\begin{center}
\begin{tabular}{ |c|c|c|c| } 
    \hline
    \textbf{Component} & \textbf{\ac{DA}} & \textbf{\ac{SA}} & \textbf{\ac{FH}} \\ 
    \hline
    \ac{LNA} & $N_{BS}$ & $N_{BS}$ & $N_{BS}$ \\
    \ac{PS} & 0 & $N_{BS}$ & $N_{BS}N_{RF}$ \\ 
    Mixers & $N_{BS}$ & $N_{RF}$ & $N_{RF}$ \\
    \ac{LO} & $N_{BS}$ & $N_{RF}$ & $N_{RF}$ \\
    \ac{VGA} & $N_{BS}$ & $N_{RF}$ & $N_{RF}$ \\
    \ac{ADC}s & $N_{BS}$ & $N_{RF}$ & $N_{RF}$ \\
    \hline
\end{tabular}
\label{table:components}
\end{center}
\vspace{-5mm}
\end{table}

\section{Analysis Approach}
\label{sec:analysis}
To capture the power consumption and communications capabilities of each of these array designs, we use \ac{SE} and \ac{EE} as system metrics. \ac{SE} is commonly used as a system requirement, defined as
\begin{align}
    S = \frac{1}{K} \textstyle\sum_{k = 1}^K \textstyle\sum_{u = 1}^U \log_2\left(1 + \zeta_{u,k}\right)
    \label{eq:se}
\end{align}
where $K$ is the number of subcarriers and $\zeta_{u,k}$ is the \ac{SINR} for user $u$ at subcarrier $k$ after MIMO combining. \eqref{eq:se} computes the \ac{SE} from capacities averaged over subcarriers, but summed over all users. As in \cite{Abbas2017_mmwRx}, the \ac{EE} is defined by the data rate achieved per unit power consumed: $E = \frac{S\cdot B}{P}$ where $B$ is the total system bandwidth and $P$ is the total receiver power consumption. Since \ac{EE} depends on \ac{SE}, the tradeoff was computed in three steps: 1) simulate the system \ac{SE}; 2) calculate the total circuit power from the circuits literature survey; and 3) assess the \ac{EE} using the assumed $B$ and computed $S$ and $P$.

Wideband \ac{MC} simulations implemented MIMO-OFDM to compute the \ac{SE}. Each user is assumed to have an analog phased array \ac{TX} with $N_U$ antennas. By concatenating all $U$ user \ac{TX} arrays and their channels, we modeled the system at subcarrier $k$ in \eqref{eq:system}, where $\widetilde{\mathbf{Y}}$ represents the concatenated received streams for all users, $\mathbf{W}_D \in \mathbb{C}^{N_{RF} \times U}$ is the digital combiner, $\mathbf{W}_{RF} \in \mathbb{C}^{N_{BS} \times N_{RF}}$ is the analog combiner (an identity matrix for the \ac{DA}), $\mathbf{V}_{RF} \in \mathbb{C}^{UN_U \times U}$ is the concatenated analog precoding for all users, and $\mathbf{Z}$ is the Gaussian noise at the antennas.
\begin{align}
    \widetilde{\mathbf{Y}} [k] = \mathbf{W}^H_D[k]  \mathbf{W}^H_{RF} \mathbf{H}[k] \left(\mathbf{V}_{RF} \mathbf{S}[k] + \mathbf{Z}[k]\right) 
    \label{eq:system}
\end{align}

The precoder and combiner matrices were computed using the algorithms detailed in \cite{Sohrabi2016_combiner} and \cite{Sohrabi2017_combiner}, with the multiple users acting as a distributed sub-array \ac{TX}. Due to the \ac{PS} constraints, each entry of $\mathbf{W}_{RF}$ and $\mathbf{V}_{RF}$ must be unit magnitude. Additionally, the array architectures require $\mathbf{W}_{RF}$ for the \ac{SA} and $\mathbf{V}_{RF}$ to be block diagonal matrices. Thus, $\mathbf{W}_{RF}$ and $\mathbf{V}_{RF}$ required an iterative algorithm due to these hardware constraints on the matrix solutions and the wideband \ac{sub-THz} channel, while $\mathbf{W}_D$ was computed with the MMSE solution. As one of the few \ac{sub-THz} channel models based on experimental data, the simulations used NYUSIM 3.0 \cite{Ju2021_nyusim3}, a statistical channel model based on indoor measurements at 140 GHz. We used 10 \ac{MC} trials to average the \ac{SE} over many channel realizations, totaling 80 independent user channels. 

Each \ac{MC} trial simulated 1000 symbols to estimate the stream \ac{SINR}. Symbols for each subcarrier were modeled using zero-mean, complex Gaussian random numbers with variance defined by the signal power. We used Gaussian symbols to represent a generalized amplitude and phase modulation. Simulations estimated the user SINR for each subcarrier in \eqref{eq:sinr}, applying the MMSE solution $\widehat{g}_{u,k} = \boldsymbol{s}_{u,k}^\dagger \boldsymbol{y}_{u,k}$ for the output signal gain $g_{u,k}$ using known transmitted symbols $\mathbf{s}_{u,k}$ and received symbols $\mathbf{y}_{u,k}$. Table \ref{table:sims} lists the other critical system assumptions used in the simulations.
\begin{align}
    \label{eq:sinr}
    \zeta_{u,k} = \textstyle \frac{\norm{\widehat{g}_u\boldsymbol{s}_u}^2}{\widehat{\sigma}_{int}^2}, \quad \widehat{\sigma}_{int}^2 = \mathbb{E}\norm{\boldsymbol{y}_u - \widehat{g}_u \boldsymbol{s}_u}^2
\end{align}

\begin{table}[htbp]
\caption{Simulation settings and assumptions}
\vspace{-3mm}
\begin{center}
\setlength{\arrayrulewidth}{0.22mm}
\begin{tabular}{ |g|c||c|c| } 
    \hline
    Channel scenario & \multicolumn{3}{c|}{Indoor office \ac{LOS}} \\ 
    \hline
     & \multicolumn{3}{c|}{$[(16\times4),(32\times4),(24\times8),(32\times8),$} \\ \arrayrulecolor{lightgray} \cline{1-1}\cline{1-1}\cline{1-1} \arrayrulecolor{black} 
    \multirow{-2}{*}{$N_{BS}$} & \multicolumn{3}{c|}{$(48\times8),(32\times16),(48\times 16),(64\times16)]$} \\ 
    \hline 
    $N_U$ & $16 \times 4$ & \cellcolor{lightgray} $U = N_{RF}$ & 8 \\
    \hline
    Number of Subcarriers & 256 & \cellcolor{lightgray} $B$ & 800 MHz \\
    \hline
\end{tabular}

\label{table:sims}
\end{center}
\vspace{-5mm}
\end{table}

\section{Sub-THz Component Power and Survey}
\label{sec:litSurvey}
The total \ac{RX} power was modeled as a sum of component powers determined from our literature survey \cite{ref_litsurvey}. Explicitly, the power is computed in \eqref{eq:powertotal}, with the component powers described in the remainder of this section.
\begin{align}
    \label{eq:powertotal}
    P = \left(P_{LNA} \cdot N_{BS}\right) + \left(P_{RF} \cdot N_{RF}\right) + P_{DSP} \\
    \label{eq:powerRF}
    P_{RF} = P_{LO} + P_{Mixer}+ P_{VGA} + P_{ADC}
\end{align} 

\ac{LNA} power was computed based on a \ac{FoM} from our literature survey, as in \eqref{eq:powerLNA}. \cite{ref_lna} showed the highest \ac{FoM} in the survey, with $FoM_{LNA} = 1.84$ mW\textsuperscript{-1}. We assumed a \ac{NF} of 10 dB ($F_{LNA}=10$) and gain $G_{LNA} = 26$ dB, as in the selected \ac{LNA} design, translating to a unit power of 24 mW. The gain was fixed to determine a realistic \ac{LNA} power, though higher power \ac{LNA}s with larger gains may be needed to meet implementation specific \ac{RX} \ac{NF} requirements. The remainder of the gain required in the signal path was generated in the \ac{BB} \ac{VGA}s.

\begin{align}
    \label{eq:powerLNA}
    P_{LNA} = 10^{G_{LNA}/10}\left(FoM_{LNA}(F_{LNA} - 1)\right)^{-1}
\end{align}

The \ac{sub-THz} survey included both active and passive mixer designs. As in \cite{Skrimponis2020_subthzMobileRx}, we selected a passive mixer ($P_{Mixer} = 0$) to minimize DC power consumption, although active mixers may be more suitable depending on implementation specific \ac{ADC}, \ac{VGA}, \ac{LO}, and \ac{NF} requirements. With a conversion loss of $I_M = 9.8$ dB, the lowest among passive mixers, \cite{ref_mixer} minimized the \ac{VGA} gain required. The \ac{LO} in \cite{ref_lo} was then selected for having the highest DC-to-RF efficiency and sufficient RF output for the mixer. If signal distribution losses are low, the total power potentially could be reduced by using one \ac{LO} to drive two mixers, since the \ac{LO} in \cite{ref_lo} outputs 3.3 dB more RF power than the mixer in \cite{ref_mixer} requires. For simplicity, we assumed that each mixer required its own \ac{LO}.

\ac{PS}, splitter, and combiner losses affect the \ac{VGA} power by determining the maximum gain required. \ac{PS} power and \ac{IL} vary based on the required resolution; active \ac{PS}'s generally support higher resolutions than passive designs but require DC power. In our power analysis, we considered two \ac{PS} cases: passive \ac{PS}'s based on \cite{ref_psPassive} with an \ac{IL} of $I_P = 6$ dB and active \ac{PS}'s based on \cite{ref_psActive} requiring 30 mW of power and $I_P = 5.8$ dB \ac{IL}. As shown in the literature survey, splitters and combiners at \ac{sub-THz} have significant insertion loss. Based on \cite{ref_combiner}, we assumed an \ac{IL} of $I_S = I_C = 1.3$ dB for up to 8 output traces for splitters and 8 input traces for combiners. The total distribution loss was computed based on the number of series splitters and combiners required to support enough RF signal paths in each hybrid architecture.

The \ac{BB} \ac{VGA} was modeled to provide the required input voltage range for the \ac{ADC}s. Based on the survey, \ac{VGA}s tend to use nearly the same power within their designed gain range. Using \cite{ref_bbamp}, we assumed a unit power of $P_{v} = 10.8$ mW for each unit of $G_{v} = 20$ dB gain required. We selected the total \ac{BB} gain to amplify the weakest input signals, assuming a minimum per-antenna SNR of 0 dB ($\zeta = 1$) and thermal noise at $T = 300$K, to the \ac{ADC}'s peak-to-peak input swing $V_a$ (V). \eqref{eq:powerVGA} shows the total power for each set of \ac{VGA}s to meet the required total gain in \eqref{eq:gainVGA}, where $P_s = \zeta P_n$ and $P_n = k_B T B$ are the minimum per-antenna input signal and noise power (W) respectively and $k_B$ is the Boltzmann constant.
\begin{align}
    \label{eq:powerVGA}
    P_{VGA} &= P_{v} \cdot \ceil[\big]{\textstyle\frac{G_{VGA}}{G_{v}}} \\
    \label{eq:gainVGA}
    G_{VGA} &= 10\log_{10}\left(\textstyle\frac{V_a^2}{8 R_i}\frac{1}{P_s + F_{LNA} P_n}\right) - G_{LNA} \nonumber \\ &\quad + I_S + I_P + I_C + I_M
\end{align}

\ac{ADC} and \ac{DSP} power are both computed using \ac{FoM}s. The \ac{ADC} power was computed in \eqref{eq:powerADC}, with the Walden \ac{FoM}, sampling frequency $F_s = 2B$, and an \ac{ADC} resolution of $N_b$ bits. We assumed $FoM_{ADC} = 40$ fJ/step/Hz, the median \ac{FoM} of designs surveyed in \cite{Murmann2021_adcsurvey} over the last 6 years, and $V_a = 0.5$ V from \cite{ref_adc}. \eqref{eq:powerDSP} computes the \ac{DSP} power with $FoM_{DSP} = 13$ GOPS/mW, where $U \cdot \left(2N_{RF} - 1\right)$ represents the number of operations to apply the digital combiner. Although the $\mathbf{W}_{D}[k]$ is applied to every subcarrier, the total number of operations does not scale with $K$, as the results are only computed for each OFDM symbol instead of each sample. The \ac{DSP} \ac{FoM} was selected from \cite{Yuan2014_dspfom} for a 40 nm CMOS radio processor.
\begin{align}
    \label{eq:powerADC}
    P_{ADC} = FoM_{ADC} \cdot F_s \cdot 2^{N_{b}} = FoM_{ADC} \cdot B \cdot 2^{N_{b} + 1} \\
    \label{eq:powerDSP}
    P_{DSP} = U \cdot \left(2N_{RF} - 1\right) \cdot B \cdot FoM_{DSP}^{-1}
\end{align}

\section{Tradeoff Results}
\label{sec:results}

\begin{figure}
    \centering
    \includegraphics[width=0.95\columnwidth,trim=5 0 15 2,clip]{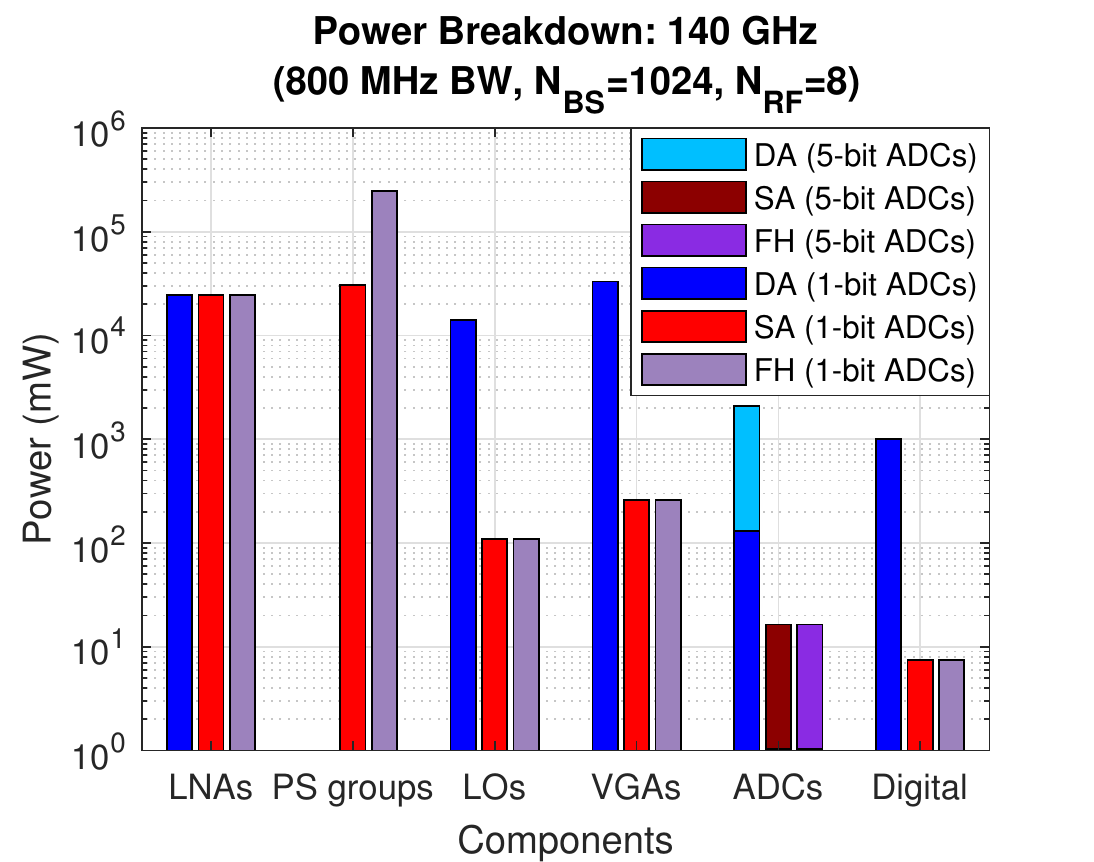}
    \caption{Breakdown of \ac{RX} power consumption by component}
    \label{fig:powerBreak}
    \vspace{-4mm}
\end{figure}

\begin{figure}
    \centering
    \includegraphics[width=0.95\columnwidth,trim=5 0 15 2,clip]{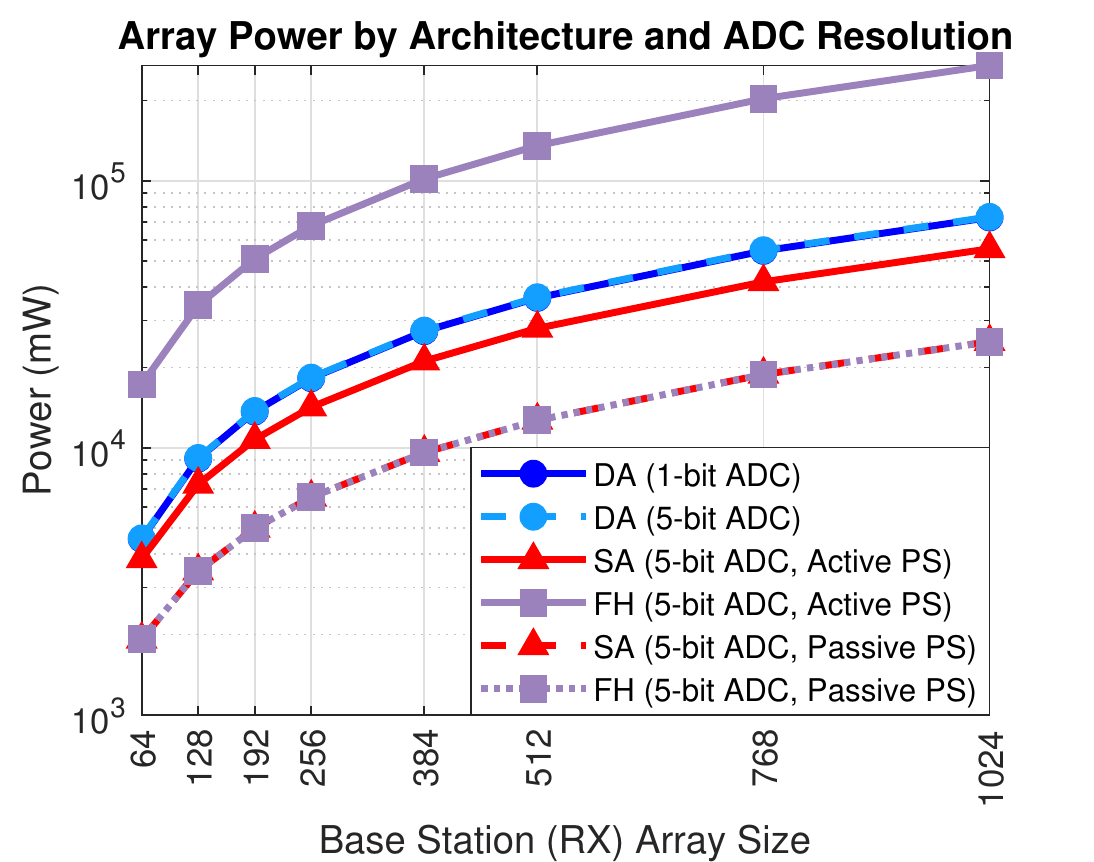}
    \caption{Total power for primary \ac{RX} configurations}
    \label{fig:powerTotal}
    \vspace{-4mm}
\end{figure}

The total \ac{RX} power consumption depends primarily on the architecture and the \ac{PS}'s. A breakdown of the power consumption by component is shown in Fig. \ref{fig:powerBreak}. Since the active \ac{PS}'s have nearly the same \ac{IL} as the passive \ac{PS}'s, this graph can illustrate power for both cases. If the \ac{PS} group bars are ignored for the passive \ac{PS} case, Fig. \ref{fig:powerBreak} shows that the \ac{DA} uses much more power than other architectures in nearly all components. However, for the active \ac{PS} case, \ac{PS}'s consume the most power of any component in the \ac{SA} and \ac{FH} due to the sheer number of devices. Fig. \ref{fig:powerTotal} compares the total power of several design combinations and illustrates the impact of \ac{PS}'s clearly. Requiring active \ac{PS}'s increases the \ac{FH}'s power by an order-of-magnitude and increases the \ac{SA}'s power to be nearly the same as that of the \ac{DA}. 

Although higher \ac{ADC} resolution exponentially increases \ac{ADC} power, high resolution does not significantly affect total power in any of the three architectures. Fig. \ref{fig:powerBreak} demonstrates that the RF and \ac{BB} components consume an order-of-magnitude or more power than 5-bit \ac{ADC}s, while Fig. \ref{fig:powerTotal} illustrates the minimal impact \ac{ADC} resolution has on the total \ac{RX} power. This conclusion is consistent with \cite{Skrimponis2020_subthzMobileRx}, although different assumptions on gain distribution shift the total power consumption amongst the \ac{BB} and RF components.

\begin{figure}
    \centering
    \includegraphics[width=\columnwidth,trim=5 0 15 2,clip]{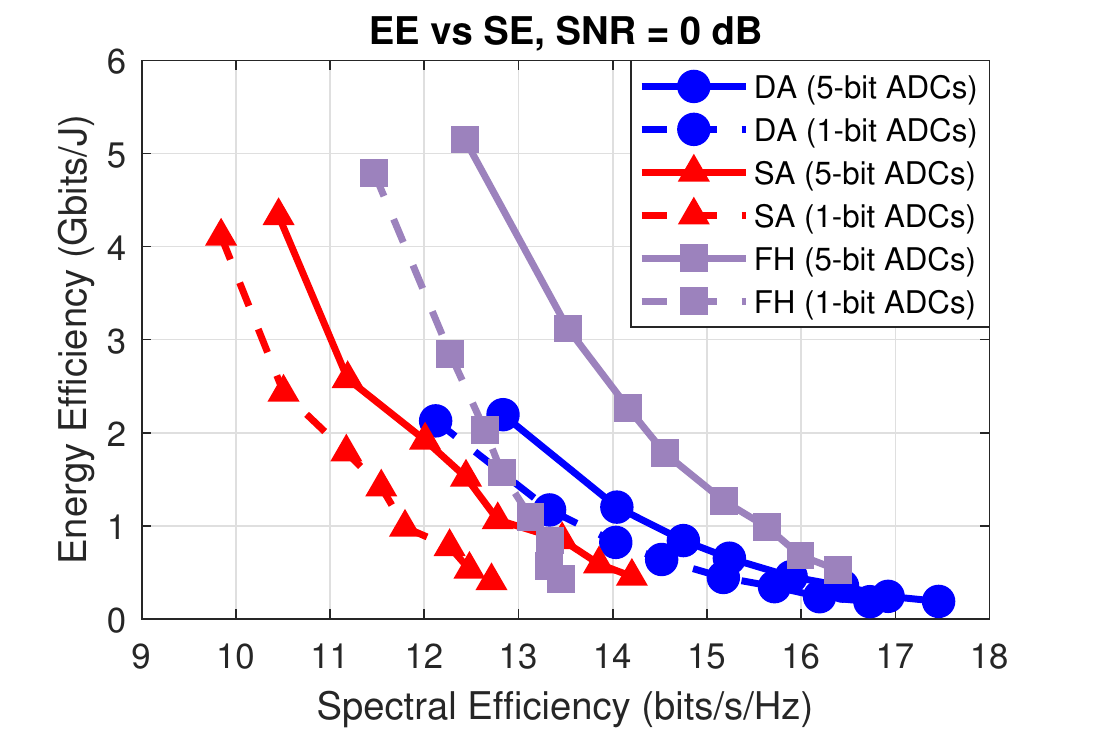}
    \caption{\ac{EE} vs \ac{SE} for passive \ac{PS} and low SNR}
    \label{fig:EEvsSE_passive_0dB}
    \vspace{-5mm}
\end{figure}

The \ac{SE} and \ac{EE} tradeoffs summarize the reduced power efficiency resulting from higher communication capability requirements, combining the impacts of component power, architecture, and assumed SNR. Note that each point on a given \ac{EE} vs \ac{SE} curve represents a different array size; larger arrays generally have higher \ac{SE} and lower \ac{EE}, corresponding to the bottom right of the graphs. Fig. \ref{fig:EEvsSE_passive_0dB} provides the worst scenario for the \ac{DA} with low per-antenna SNR (0 dB) and passive \ac{PS}'s. The low SNR limited the array's maximum capacity, inhibiting the \ac{DA}'s ability to take advantage of more combiner parameters and resulting in similar \ac{SE} performance between the \ac{DA} and the \ac{FH}. In this case, the \ac{FH} with 5-bit \ac{ADC}s provided the best \ac{EE} vs \ac{SE} tradeoff, since passive \ac{PS}'s allowed the \ac{SA} and the \ac{FH} to utilize much less power than the \ac{DA}. Only in cases where the maximum \ac{SE} is required, around 1 bits/s/Hz higher than the \ac{FH}'s maximum, is the \ac{DA} the only option and thus the most efficient.

\begin{figure}
    \centering
    \includegraphics[width=\columnwidth,trim=5 0 15 2,clip]{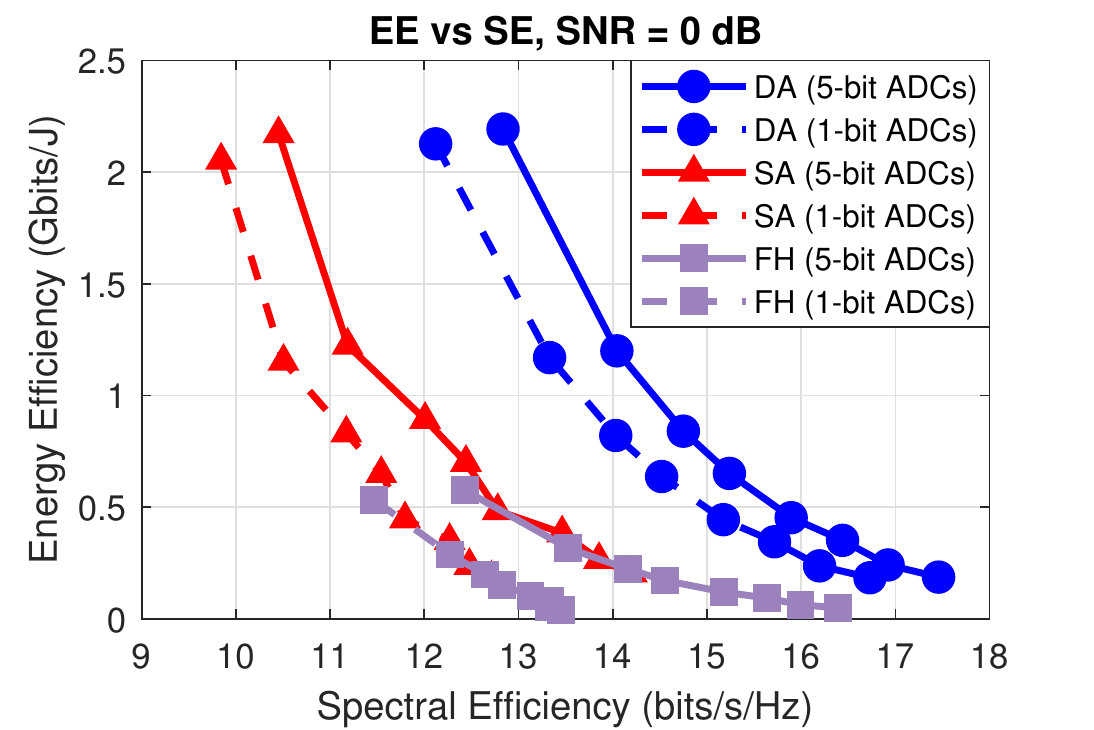}
    \caption{\ac{EE} vs \ac{SE} for active \ac{PS} and low SNR}
    \label{fig:EEvsSE_active_0dB}
    \vspace{-5mm}
\end{figure}

\begin{figure}
    \centering
    \includegraphics[width=\columnwidth,trim=5 0 15 2,clip]{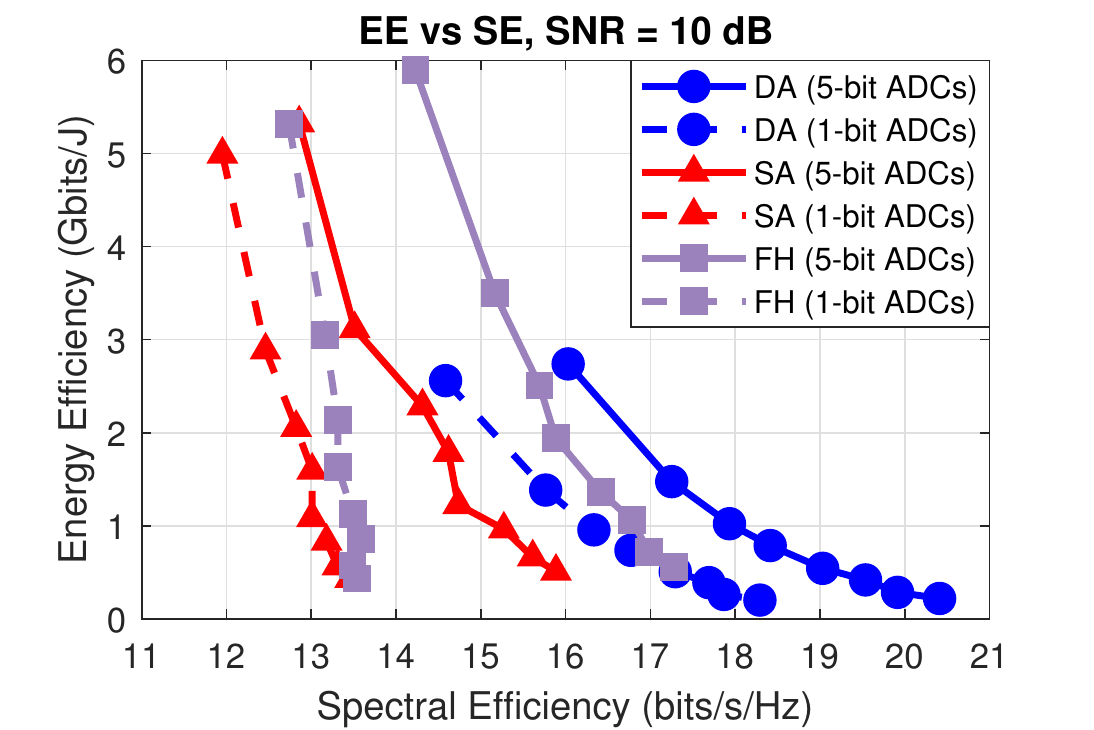}
    \caption{\ac{EE} vs \ac{SE} for passive \ac{PS} and moderate SNR}
    \label{fig:EEvsSE_passive_10dB}
    \vspace{-5mm}
\end{figure}

Requiring active \ac{PS}'s or higher SNRs made the \ac{DA} more competitive. Fig. \ref{fig:EEvsSE_active_0dB} shows the active \ac{PS} case with 0 dB SNR. The huge increase in power for the \ac{SA} and \ac{FH} with active \ac{PS}'s reduced the \ac{EE} of the \ac{SA} by half and the \ac{EE} of the \ac{FH} by nearly 10 times. Thus, the \ac{DA} became more efficient than the hybrid designs at high \ac{SE}s and the \ac{FH} was not viable under any \ac{SE} requirement. Fig. \ref{fig:EEvsSE_passive_10dB} demonstrates the benefit of higher per-antenna SNRs for the \ac{DA}. While the hybrid arrays saw small increases in \ac{SE}, the \ac{DA} achieved nearly 3 bits/s/Hz improvement. Thus, at higher SNRs, the \ac{DA} with 5-bit ADCs became the most efficient option for high \ac{SE} requirements.

\section{Conclusion and Future Work}
\label{sec:conclusion}
This paper presents realistic power consumption and tradeoffs in \ac{EE} for required \ac{SE}. \ac{MC} communication simulations and a survey of state-of-the-art \ac{sub-THz} circuit components provides a realistic analysis of \ac{BS} \ac{RX} capacity and power consumption. Our results show that the \ac{DA} can be the most efficient architecture, especially in high SNRs, when active \ac{PS}'s must be used, or when high \ac{SE} is required.
Sub-THz \ac{TX} considerations, realistic \ac{sub-THz} cellular link budgets, the impact of \ac{PS} quantization on hybrid combining, and additional channel environments are left as future work for \ac{sub-THz} \ac{BS} analysis. Low power \ac{sub-THz} \ac{LNA}s, \ac{LO}s, and \ac{PS}'s could significantly improve the efficiency of future array designs.

\pagebreak



\bibliographystyle{IEEEtran}
\bibliography{IEEEabrv,references}

\end{document}